# Pathways of organic micropollutants degradation in atmospheric pressure plasma processing – A review

Barbara Topolovec [a,b]. Nikola Škoro [c]. Nevena Puač [c]. Mira Petrovic [a,d] *

[a] Catalan Institute for Water Research (ICRA), Emili Grahit 101, 17003, Girona, Spain

[b] University of Girona, Girona, Spain

[c] Institute of Physics, University of Belgrade, Pregrevica 118, 11080 Belgrade, Serbia

[d] Catalan Institution for Research and Advanced Studies (ICREA), Passeig Lluis Companys 23, 08010 Barcelona, Spain

*Corresponding author

Mira Petrovic. Catalan Institute for Water Research (ICRA), Emili Grahit 101 17003, Girona, Spain; Catalan Institution for Research and Advanced Studies (ICREA), 08010; Phone: (+34) 972 18 33 80; E-mail: mpetrovic@icra.cat

Abstract

Concern of toxic compounds and their, potentially more harmful degradation products, present in aquatic environment alarmed scientific community and research on the development of novel technologies for wastewater treatment had become of great interest. Up to this date, many papers pointed out the challenges and limitations of conventional wastewater treatment and of some advanced oxidation processes. Advanced technologies based on the use of non-equilibrium or non-thermal plasma had been recognized as a possible solution for, not only degradation, but for complete removal of recalcitrant organic micropollutants. While previous review papers have been focused on plasma physics and chemistry of different types of discharges for few organic micropollutants, this paper brings comprehensive review of current knowledge on the chemistry and degradation pathways by using different non-thermal plasma types for several micropollutants' classes, such as pharmaceuticals, perfluorinated compounds, pesticides, phenols and dyes and points out some major research gaps.

1. Introduction

During the last decades, an increasing number of emerging contaminants (EC) have been detected in the aquatic environment. Among them, the main concern and awareness is about the presence of organic micropollutants (OMPs). OMPs represents wide group of chemical compounds, including, but not limited to, pharmaceuticals, pesticides, personal care products, different household, and industrial chemicals, which can be found in wastewaters of urban, industrial, or agricultural origin. OMPs in water bodies are known to cause harmful effects on aquatic organisms and ultimately on humans (EEA (European Environment Agency), 2011; Rozas et al., 2016; Verlicchi et al., 2012). Various studies report that some OMPs are bioaccumulative and can cause growth inhibition and endocrine disruption in aquatic organisms (Caballero-Gallardo et al., 2016; Santos et al., 2010; Tijani et al., 2014).

Abundant research conducted in the past showed that conventional biological treatment, adopted at majority of wastewater treatment plants (WWTPs) is only partially able to mitigate OMPs contamination with highly variable performance (Falås et al., 2016; Gros et al., 2010). Poor elimination of more hydrophilic and poorly-to-moderately biodegradable OMPs is well documented (Luo et al., 2014). Advanced technologies to remove OMPs from the wastewater are available, but in many cases, they are too costly and/or energy unsustainable. In previous studies, different advanced oxidation processes (AOPs), such as ultraviolet (UV)/hydrogen peroxide ($H_2O_2$) treatment, ozonation ($O_3$), photo-Fenton have been recognized as successful for wastewater treatment with the ability to rapidly and efficiently degrade many different types of organic compounds (Giannakis et al., 2015; Kudlek, 2018; Wols & Hofman-Caris, 2012).

One of the novel highly-promising AOPs is based on the use of non-equilibrium or non-thermal plasma (NTP) which produce nitrogen-containing species, hydroxyl radical, hydroperoxyl radical, atomic hydrogen and oxygen, as well as other radicals and active chemical species that are generally required for an effective water treatment (Ghime & Ghosh, 2020; Miklos et al., 2018; Ribeiro et al., 2015). However, little is known about the feasibility of plasma treatment for the removal of more recalcitrant compounds from wastewater (i.e.OMPs) or regarding their degradation kinetics, as well as degradation by-products generated and their toxicity.

In previous reviews on water treatment which involve plasma technology, the focus was on plasma physics and chemistry of different types of discharges and modeling of plasma interaction with the liquid (Foster, 2017; Foster et al., 2012; Hijosa-Valsero et al., 2014; Jiang et al., 2014; Lindsay et al., 2015; Parvulescu et al., 2012; Reuter et al., 2018). This review summarizes the current knowledge on the chemistry and pathways of degradation of OMPs by using different NTP types.

2. Fundamentals of non-equilibrium plasma for water treatment

2.1 Plasma classification and technological aspects

Plasma is fully or partially ionized gas and is frequently called the fourth state of matter. Generally speaking, by applying energy to a gas, some amount of excited species and ions and electrons are produced in frequent collisions between the existing free electrons and neutral species eventually leading to the formation of plasma. Energy input required for the creation of plasma in majority of cases is realized by imposing sufficiently high electric field. Although it contains free charge carriers, from a macroscopic point of view, plasma is electrically neutral. In thermodynamic term plasma species (electrons, ions, and neutrals) can be characterized by their temperatures and in the respect to this description plasma can be classified into two categories: equilibrium or thermal and non-equilibrium plasma. In thermal plasmas characteristic temperatures of electrons and heavy particles (atoms, molecules, ions) are equal high, with temperature values of $10^3$ K and above (Lieberman & Lichtenberg, 1994). Strictly speaking, at terrestrial conditions (i.e., in applications) these kinds of plasma are only in local thermodynamic equilibrium since the temperatures of all plasma species are the same only in a limited space volume. When comes to applications, thermal plasmas are realized as arcs and torches and typically used for applications where high temperatures are required, such as for cutting, spraying, welding or, as in the analytical devices, for the evaporation of an analyte material (Tendero et al., 2006).

In non-equilibrium plasmas the temperature of the electrons is much higher than the temperatures of heavy particles. Depending on the type of the plasma, temperatures of the heavy particles can be few hundred degrees or much below and close to the room temperature. Thus, these plasmas are also referred as non-thermal plasma (NTP) or cold plasma. Due to pronounced non-equilibrium of particle energies, mainly electrons have sufficient energy to collide with the background gas particles and provide multiplication of charges necessary to sustain the plasma. Apart from production of charged particles, excited species (especially metastables), fast neutrals, radicals, photons are also produced in collisions. In comparison to thermal plasmas, non-equilibrium plasma is characterized by lower energy densities, have lower degree of ionization and consequently lower particle densities (Lieberman & Lichtenberg, 1994). Nevertheless, the fact that these plasmas operate at temperatures which are equal or close to room temperatures, allow their application in many different aspects (Becker et al., 2004; Makabe & Petrovic, 2014).

Non-equilibrium plasmas working at atmospheric pressure have an advantage compared to their low-pressure counterparts when comes to applications where samples cannot withstand exposures to pressures below atmospheric. Thus, many applications of plasma in fields of biology and medicine have been driving the development of atmospheric pressure non-equilibrium plasmas in the last 30 years (Fridman et al., 2008; Laroussi et al., 2017; Machala et al., 2012; Miletić et al., n.d.). Several types of electrical discharges at atmospheric pressure, such as corona, dielectric barrier discharge (DBD), gliding arc, glow discharge and streamer discharge, have been employed depending on particular treatment demands. Albeit the limited active volume of these plasmas versatility and success of their biomedical applications lies in production of suitable chemical species and their interaction with living cells (Lu et al., 2016; Privat-Maldonado et al., 2019; Tomić et al., 2021).

In recent years, the applications of non-equilibrium plasmas are expanded to the wastewater treatment. According to recent papers, applications of NTP for wastewater treatment are becoming more attractive and are recognized as technological solution for EC which are highly recalcitrant to conventional wastewater treatment technologies. Reactive chemical species created in the plasma have already proven their capability to decontaminate water by pulverizing molecules of the pollutant in various chemical reactions (Foster, 2017; Magureanu et al., 2018). However, before reaching the auspicious situation where plasma technology is used for wastewater treatment several major obstacles should be taken. The first one is related to determination of proper parameter related to the power consumption in plasma processes that will adhere to existing wastewater technology standards and enable comparison with conventional methods. Diversity of experimental setups for water treatment, which sometimes include additional power consumption (e.g., for sample recirculation), and plasma devices that require different approaches in power measurements make this a demanding task. Comparison of reported data in review papers shows the absence of standard for reporting of the treatment conditions (Jiang et al., 2014; Magureanu et al., 2018; Malik, 2010; Nzeribe et al., 2019). Therefore, a paper describing wastewater plasma treatment should contain the minimum information on: significant dimensions of the experimental setup, type of the discharge used, working gas flow (if present), power supply type, electrical parameters including the power delivered to plasma, volume of the treated sample (and the flow rate if in flowing regime), initial and final concentrations of the pollutant and duration of the treatment. Providing a detailed account on experimental conditions will enable analysis of and comparison of plasma treatments and can be a first step towards defining appropriate power-efficiency parameter.

Another difficulty for application of NTP as technology for water treatment come from the rather small volume of the active plasma region which, in turn, limit the out flux of reactive species created in the plasma. Since plasma is created at atmospheric pressure, mean free paths (the mean distance between two collisions) of all particles are in the order of μm. Thus, to establish an effective interaction between the plasma and the water they should be in close contact while an increase in the transfer of reactive species requires increasing the effective interaction area. In the existing experimental setups, there are two different approaches – one option is to establish plasma inside the water while another way is to create plasma in the gas phase adjacent to the liquid sample.

In that view, we can divide atmospheric pressure non-equilibrium plasmas that can be used in wastewater treatment as discharges in liquids and in contact with liquids.

### 2.2. Non-equilibrium plasmas in liquids

This group of high non-equilibrium plasmas refers to the discharges where electrodes are completely immersed in liquid. There are two sub-groups:

 i.   discharges in liquids where no gas phase plasma is involved
 ii.  discharges in bubbles inside liquids.

The major challenges in these types of discharges are how to achieve a breakdown without subsequent transition into the thermal plasma, the appropriate power supply, the optimal geometry of the electrodes and their etching during the operation, heating of the liquid, the complexity of the chemical composition of the liquid etc. Until now several review papers have been published on the topics related to the discharges in liquids (Peter Bruggeman & Leys, 2009; Graham & Stalder, 2011; Locke et al., 2006; Malik et al., 2001).

The discharges with no gas phase involved can be created in the pin-to-pin, pin-to-plate or plate-to-plate electrode system with the pulsed (usually nanosecond) power supplies (Akiyama, 2000; An et al., 2007; Locke & Thagard, 2012; Petr Lukes et al., 2005; Schaper et al., 2011; Schoenbach et al., 2008; Šunka, 2001). There are several important parameters that will influence the formation and propagation of the discharge like the energy deposited in the pulse, frequency and duty cycle, inter-electrode distance, conductivity of the liquid. The discharges are streamer-like in case of non-equilibrium plasmas and the created streamers usually dissipate before reaching the grounded electrode.

The second sub-type of the discharges in water involves creation of bubbles and in this case, we are dealing with gash phase surface plasmas (inside bubble) with the liquid electrode. The bubbles can be created by bubbling systems, heat wave, in capillary systems etc. (Bruggeman et al., 2008; Gershman et al., 2007). The bubbles do not need to be introduced into the liquid (for example by bubbling system) in order to have this type of discharge. The appearance of the micro-bubbles can be related to the breakdown of the discharge where the initial Joule heating of the liquid leads to their creation (Ceccato et al., 2010; Korobeinikov & Yanshin, 1983).

### 2.3. Non-equilibrium plasmas in contact with liquid surfaces

In this case the powered electrode is not in direct contact with liquid which serves here as a grounded electrode. There are various geometries that are used in these type of discharges: pin-jet type, DBD-jet type, planar strip electrodes that can be in direct contact with surrounding gas or immersed in dielectric etc. (Boselli et al., 2014; Puač et al., 2012; Šimor et al., 2002; Škoro et al., 2018). The used power supplies range from nano-pulsed to a sine wave excitation voltage and the applied voltages depend on the excitation frequency, working gas used, electrode geometry and electrode gap. The applied frequencies can range from several Hz up to the MHz region. The discharge is created in the gas phase and the resulted chemistry inside the liquid is the result of gas phase and gas/liquid interface chemistry (Bruggeman et al., 2016; Parvulescu et al., 2012; Sunka et al., 1999). The gas temperature stays in the range 300 K-400 K and usually the energy per pulse is of order of µJ. The electron temperature stays in the range of 1-3 eV and electron concentration between $10^{19}$-$10^{21}$ m$^{-3}$.

Table 1. Different plasma configurations used treatment of organic micropollutants

| OMPs | Plasma system/plasma type and configuration/working gas | Solution type/volume treated/treatment time/initial concentration | Degradation | Reference |
|---|---|---|---|---|
| Pharmaceuticals | | | | |
| Norfloxacin | DBD reactor/ discharge above liquid / Oxygen ($O_2$), nitrogen($N_2$) or air | Ultrapure aqueous solution 3 ml 2 min 200 mg/L | $O_2$/air – 98% $N_2$ – 50% | (Zhang et al., 2018) |
| Ofloxacin (OFX) and ciprofloxacin (CFX) | DBD reactor / discharge above liquid / air | Matrix solution 25 ml 5 – 25 min 10 mg/L | 66% CFX 72% OFX | (Sarangapani et al., 2019) |
| Paracetamol, Caffeine, Ceftriaxone | DBD reactor / tilted cylindrical configuration, discharge above liquid / air or $O_2$ | Ultrapure aqueous solution 100 ml 10 – 30 min 25 mg/L (parac.); 50 mg/L (caff.); 5 mg/L (ceft.) | >80% (caff.) 89% (parac.) >80% (ceft.) | (Iervolino et al., 2019) |
| Amoxicillin, Oxacillin, Ampicillin | DBD reactor / coaxial configuration, thin falling water film / $O_2$ | Tap water 200 ml 10 – 30 min 100 mg/L | Degradation occurred but percentage not specified | (Magureanu et al., 2011) |
| Ibuprofen (IBP) | DBD reactor / discharge above falling water film / air | Ultrapure aqueous solution 350 ml 15 min 60 mg/L | 85% | (Marković et al., 2015) |
| Diclofenac (DCF), IBP | DBD reactor / planar configuration above falling water film / argon (Ar) | Ultrapure aqueous solution 500 ml 20 – 30 min 50 mg/L | Complete degradation occured | (Hama Aziz et al., 2017) |
| DCF, Carbamazepine (CBZ), CFX | Corona discharge / pulsed needle-to-plane continuous flow reactor / air | Distilled, lake and river water 10 – 40 ml/min 24 min 1 ml (in mix added in equal ratio) | >99% in distilled and lake water >91% in river water | (Singh, Philip, et al., 2019) |
| Paracetamol | Corona discharge / pulsed above liquid / air or $O_2$ | Ultrapure aqueous solution 40 000 ml 30 min 100 mg/L | Complete degradation with partial mineralization | (Panorel et al., 2013) |
| CBZ, diatrizoate, diazepam, DCF, IBP, 17a-ethinylestradiol, trimethoprim | Corona discharge / coaxial geometry / discharge inside water in surrounding air | Ultrapure aqueous solution 300 ml 60 min 500 mg/L | 45 – 99 % | (Banaschik et al., 2015) |
| Perfluorinated compounds | | | | |
| perfluorooctanoic acid (PFOA), perfluorooctane sulfonate (PFOS) | DC-plasma generated within gas bubbles/ $O_2$, helium (He), Ar | Ultrapure aqueous solution 50 mL 240 min 50 mg/L | 92% - decomposition ratio of fluorine ions 57% decomposition ratio of sulfate ions | (Yasuoka et al., 2010) |
| PFOA, PFOS | DC-plasma generated within gas bubbles/ $O_2$ | Ultrapure aqueous solution 20 mL PFOA: 3h 41,4 mg/L PFOS: 8h 60 mg/L | PFOA: 94,5 % defluorination ratio PFOS: 70% deflurination ratio | (Hayashi et al., 2015) |
| PFOA | DC-plasma generated within gas bubbles/$O_2$ | Ultrapure aqueous solution (single and mix with other PFAS) | Complete degradation in single solution ; 80% in mix solution | (Takeuchi et al., 2014) |

| OMPs | Plasma system/plasma type and configuration/working gas | Solution type/volume treated/treatment time/initial concentration | Degradation | Reference |
|---|---|---|---|---|
| | | 20 mL<br>150 min<br>156 µM | | |
| PFOA | Laminar jet with bubbling (LJB)/ Ar | Ultrapure aqueous solution<br>1,4 L<br>30 min<br>20µM | 90% (high removal rate process)<br>25% (high removal efficiency process) | (Stratton et al., 2017) |
| PFOA, PFOS | Discharge above the liquid with bubbling/argon | Ultrapure aqueous solution<br>1,5 L<br>30-120 min<br>8,3 mg/L | 90% of PFOA in 60 min and PFOS in 40 min | (Singh, Fernando, et al., 2019) |
| Pesticides | | | | |
| Lindane | DBD<br>a) conventional batch reactor (R1)/ He<br>b) falling water film reactor (R2) | Distilled and wastewater<br>175 mL<br>5 min<br>1 mg/ml | 87 % in R1 and 79% in R2 for distilled water<br>50% in R1 for wastewater | (Hijosa-Valsero et al., 2013) |
| Atrazine | pDBD, gas phase above liquid/with or without membrane/air | Water matrix<br>100 ml<br>45 min<br>30 µg/L | Without membrane 61,0 %<br>With membrane 84,7 % | (Vanraes et al., 2015) |
| | Pulsed DBD reactor with Continuous flow falling water film over activated carbon textile/air | Deionized water<br>2,5 L<br>2500 ml<br>30 min<br>200 µg/L | 93,9 % | (Wardenier, Vanraes, et al., 2019) |
| 2,4 - D | Corona discharge, needle to plane/air | Distilled, lake and river water<br>Not defined<br>24 min<br>0,2 – 2 mg/L | > 99 % in distilled and lake water<br>> 91 % in river water | (Singh, Philip, et al., 2019) |
| | Pulsed corona discharge/$O_2$ | Tap water<br>330 ml<br>60 min<br>25 mg/L | 93 % | (Bradu et al., 2017) |
| | DBD reactor with planar falling water film/ pure Ar, Ar/$O_2$ (80:20), and air | Deionized water solution<br>0,5 L<br>90 min<br>100 mg/L | Degradation occurred but percentage not specified | (Hama Aziz et al., 2018) |
| | Multiple pin-plate corona discharge/air | Aqueous solution<br>Not defined<br>6 min<br>1 mg/L | 100 % | (Singh et al., 2017) |
| Alachlor | Continuous flow pulsed DBD/$O_2$ | Deionized water solution<br>Single pass through reactor<br>30 min<br>1mg/L | 78,4 % | (Wardenier, Gorbanev, et al., 2019) |
| | DBD/plasma gas bubbled/ combined with activated carbon/air | Ultrapure aqueous solution<br>30 min<br>500 ml<br>57 µg/L | 75,1 % | (Vanraes et al., 2017) |
| Carbofuran | Corona discharge, needle to plane/air | Distilled, lake and river water<br>Not defined<br>24 min<br>0,2 – 2 mg/L | > 99 % in distilled and lake water<br>> 91 % in river water | (Singh, Philip, et al., 2019) |
| Phenols and phenolic compounds | | | | |

| OMPs | Plasma system/plasma type and configuration/working gas | Solution type/volume treated/treatment time/initial concentration | Degradation | Reference |
|---|---|---|---|---|
| 2,4 dibromophenol | DBD<br>a) conventional batch reactor (R1)/ He<br>b) falling water film reactor, surface discharge (R2) | Ultrapure aqueous solution<br>a) 4 ml<br>b) 174 ml<br>5 min<br>1 mg/L | a) 98 %<br>b) 73.5 % | (Hijosa-Valsero et al., 2013) |
| Bisphenol A (BPA) | Pulsed DBD reactor with Continuous flow falling water film over activated carbon textile/air | Synthetic aqueous solution in mix<br>2500 ml<br>30 min<br>200 µg/L | 98.8 % | (Wardenier, Vanraes, et al., 2019) |
| Phenol | DBD reactor, surface discharge/air | Ultrapure aqueous solution<br>70 ml<br>240 min<br>47,05 mg/L | Almost complete removal | (Ceriani et al., 2018) |
| | Cylindrical DBD reactor/ $O_2$ | Ultrapure aqueous solution<br>100 ml<br>15 min<br>50 mg/L | Complete mineralization after 15 min | (Iervolino et al., 2019) |
| | DBD reactor, surface discharge/air | Ultrapure aqueous solution<br>70 ml<br>120, 240 min<br>47,05 mg/L | 80 % in 4h(Ni/Cr wire)<br><br>Complete degradation in 2h (SS wire) | (Marotta et al., 2011) |
| 2,4 – dichlorophenol | DBD reactor with planar falling water film/ pure Ar, Ar/$O_2$ (80:20), and air | Ultrapure aqueous solution<br>500 ml<br>15 min<br>100 mg/L | Complete degradation | (Hama Aziz et al., 2018) |
| **Dyes** | | | | |
| Methylene blue | DBD reactor/cylindrical configuration/ air, oxygen | Ultrapure aqueous solution<br>100 mL<br>5 min;<br>10 mg/L | 96% ($O_2$)<br>76% (air) | (Iervolino et al., 2019) |
| | DC plasma, in slug flow/ gas -liquid interface with bubbles/ $O_2$, Ar, He | Synthetic wastewater<br>5 mL<br>time specified as one-time (one lap)<br>15 mg/L | 94.5 % $O_2$<br>89.3 % Ar<br>75.8 % He | (Yamada et al., 2020) |
| | Pulsed corona discharge in multiwire-plate/air | Ultrapure aqueous solution<br>35 mL<br>10 min<br>50 mg/L | Almost complete degradation(decolorization) | (Monica Magureanu et al., 2013) |
| Methyl orange | Surface glow discharge/air | Ultrapure aqueous solution<br>50 ml<br>15 min<br>10 mg/L | 93 % | (Liu et al., 2016) |
| | DBD reactor/air | Aqueous solution<br>10 mL<br>60 – 180 sec<br>50, 75, 100 mg/L | Degradation occurred but percentage not specified | (Sarangapani et al., 2017) |
| | Cylinder-like reactor, DC plasma/air | Aqueous solution<br>200 ml<br>30 min<br>20 mg/L | Degradation occurred but percentage not specified | (He et al., 2018) |
| Methyl red | DBD reactor/air | Solution in deionized water<br>25 ml<br>5 min<br>100 mg/L | Complete degradation after 5 min at 80 kV | (Pankaj et al., 2017) |

| OMPs | Plasma system/plasma type and configuration/working gas | Solution type/volume treated/treatment time/initial concentration | Degradation | Reference |
|---|---|---|---|---|
| Crystal violet | DBD reactor/air | Solution in deionized water<br>25 ml<br>5 min<br>100 mg/L | Complete degradation after 5 min at 80 kV | (Pankaj et al., 2017) |
| | DBD reactor/cylindrical configuration/air | Ultrapure aqueous solution<br>Not available<br>25 min<br>50 – 100 mg/L | > 89% | (Reddy & Subrahmanyam, 2012) |
| Acid orange 7 | Two formed modes- corona like discharge and streamer channels/air | Deionized water<br>200 ml<br>Time not defined<br>20 mg/L | Degradation occurred but percentage not specified | (Ruma et al., 2013) |

3. Application of plasma for removal of pollutants dissolved in water

3.1. Pharmaceuticals

Pharmaceuticals are OMPs which are widely investigated and the attention for further research is increasing due to the numerous reports of their presence in ground, surface and even drinking water. Despite low concentrations, the influence of continuous exposure to mixtures of different therapeutic groups has become of growing concern, not just to the aquatic species, but for human health as well (Arnold et al., 2014; Bernhardt et al., 2017). Some of these contaminants have been reported as very persistent, prone to accumulation and very hard to remove with conventional water treatment technologies. Therefore, alternative methods are being investigated and NTP, as a potent chemically active agent, has shown to be promising solution for degradation of highly resistant pharmaceuticals. Table 1 shows several plasma configurations most used for pharmaceuticals.

As it can be seen in Table 1, plasma configurations dominantly present in the papers are DBDs and corona. In the (Krause et al., 2011) rework, preceding corona discharge over water reactor (Krause et al., 2009) was further developed and used to study degradation of "CBZ" and clofibric acid. Two electrodes with dielectric barrier were installed above rotating drum, which moved a test solution as a thin water film in an atmospheric DBD system. The results showed that the degradation after 60 minutes of single-solution treatment was 94% for CBZ in ultra-pure water. Clofibric acid was degraded after 30 minutes of treatment. In a solution which contained landfill leachate, degradation after 90 minutes was 97% and 88% respectively, which showed that in the leachate the removal efficiency also depends on the type of the compound. The influence of corona discharge system on the degradation was discussed in the (Dobrin et al., 2013) for DCF. Tap water solution was treated with pulsed corona discharge above water system with addition of oxygen as working gas. Complete removal of DCF occurred after 15 minutes of treatment and after 30 minutes 50% of mineralization was achieved. The authors concluded that use of oxygen as a feeding gas improves energy efficiency as well as ozone and hydrogen peroxide formation in corona discharge that probably play important role in the degradation of the micropollutant. Similar conclusions are reached by other authors (Hama Aziz et al., 2017; Monica Magureanu et al., 2018). In a DBD reactor, with gaseous plasma-liquid system, degradation of DCF and IBP was investigated with application of different working gasses (Hama Aziz et al., 2017). The authors found that with argon, the degradation of DCF is less efficient than it was for the IBP. On the other hand, using working gas mixture of $Ar/O_2$, mineralization of only DCF was improved which implies that, even with the same DBD plasma system, the influence of gas mixture must be considered for each treated compound. Different gases and gas mixtures produce different types of radicals which effect the break of compound bonds.

Several other authors discussed the influence of different gasses and radicals on the mineralization of pharmaceuticals in their papers (Banaschik et al., 2015; Singh, Philip, et al., 2019; Wardenier, Vanraes, et al., 2019; Zhang et al., 2018) where they evaluated degradation pathways. For the degradation of norfloxacin a DBD system was used with explanation of the degradation mechanisms by (Zhang et al., 2018). Different working gasses were used in experiments: pure nitrogen or oxygen and the mixture of both for the treatment of norfloxacin solution. They discovered that different types of radical species, reactive oxygen species (ROS) and reactive nitrogen species (RNS) produced in the plasma lead to different degradation pathways. Fig.1 shows possible pathways for degradation of norfloxacin in DBD treatment with different gasses. The results showed that hydroxyl radicals may play important role in the degradation of the compound, to be precise, for the break of C-F bond in piperazine ring and formation of C-OH group. For the verification of the results, mass spectrometry analysis was used and confirmed that the products with expected masses were found. It is important to note that the solution was made with pure water and the effects may be different in the real wastewater samples.

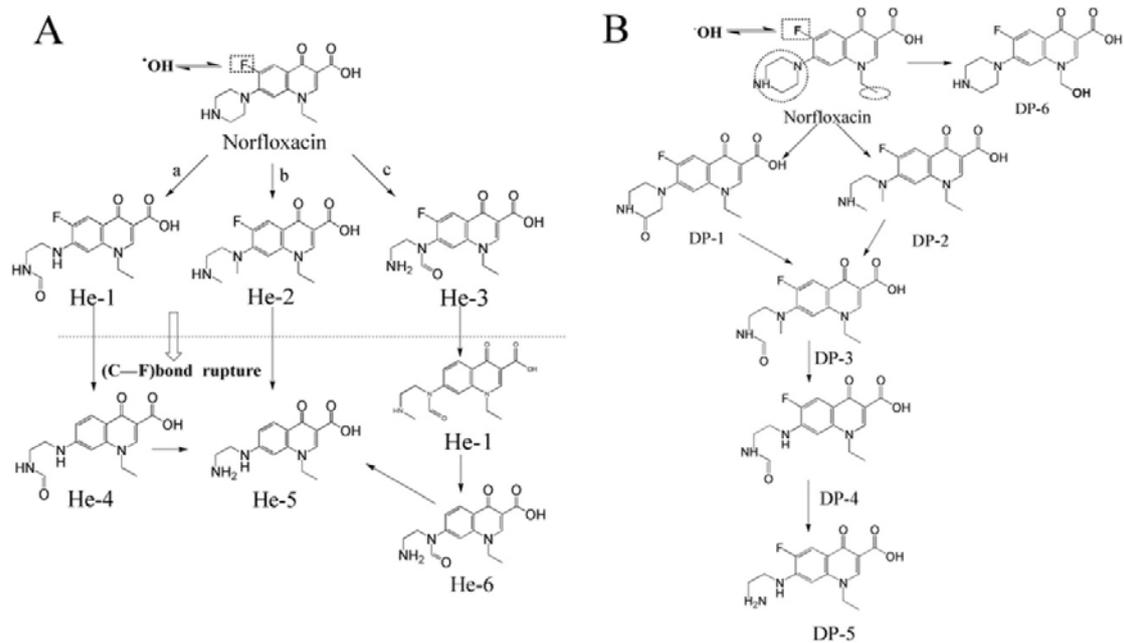

**Figure 1** Possible pathways for degradation of norfloxacin in DBD treatment: (A) treated by DBD in helium; (B) treated by DBD in air or nitrogen. Reprinted with permission from (Zhang et al., 2018). Copyright (2018) Elsevier

In the work by (Singh, Philip, et al., 2019) for three pharmaceuticals (DCF, CBZ, CPF), degradation was much faster in distilled water than in the river and lake water due to the presence of other compounds, such as carbonates and ions which can act as scavengers of *OH radicals. Also, in the presence of the nitrates, phosphates and sulfates, removal efficiency is not changed significantly which indicate that the alkalinity of the water effects on the % of degradation. Indeed, the $CO_3^{2-}$ and $HCO_3^{-}$ ions which are associated with the alkalinity are major scavengers for *OH radicals and other oxidants and in the presence of high concentrations of those ions, removal efficiency can decrease significantly. For the water pH, corona discharge in air above liquid configuration played important role since in that system, different reactive oxygen and nitrogen species are produced which can decrease the pH with the formation of H+ ions (Eqs.1-6).

$$N_2 + e \rightarrow 2N + e \tag{1}$$

$$O_2 + e \rightarrow 2O + e \tag{2}$$

$$N + O \rightarrow NO \tag{3}$$

$$NO + O \rightarrow NO_2 \tag{4}$$

$$2NO_{2(aq)} + H_2O_{(l)} \rightarrow NO_2^- + NO_3^- + 2H^+ \tag{5}$$

$$NO_{(aq)} + NO_{2(aq)} + H_2O_{(l)} \rightarrow 2NO_2^- + 2H^+ \tag{6}$$

The results given by the TOC analysis indicated that there is the presence of transformation products. The reduction in TOC was around 70%, 65% and 50% in distilled, lake and river water, respectively, but also decreasing in TOC with time indicated their mineralization. In the study by (Monica Magureanu et al., 2018) a plasma-ozonation system was employed for the treatment of IBP. It was concluded that complete removal occurred within 15-20 minutes and after 60 minutes complete mineralization was confirmed with high TOC reduction. Also, they confirmed the importance of oxygen for the degradation of OMPs and discussed importance of the discharge pulses.

In the recent work (Iervolino et al., 2019), assessed the removal of paracetamol and ceftriaxone in DBD reactor. With oxygen as a feeding gas, results showed that the complete mineralization for the paracetamol sample can be achieved after 15 minutes of treatment instead of 60 minutes in air. (Sarangapani et al., 2019) investigated influence of different radical scavengers on removal efficiency of antibiotics OFX and CFX. What was interesting is that in the presence of carbon tetrachloride ($CCl_4$), degradation of antibiotics increased. It is possible that reactions between ·H radical and $CCl_4$ may occur faster and reactions between ·H radical and hydroxyl radical is inhibited consequently giving more possibilities to hydroxyl radical and target compound (or its intermediates) reactions. Generally, when the target compound is in a matrix solution, it can affect the degradation degree induced by plasma due to other molecules and compounds that can react with oxidant radicals. Authors also suggested the degradation mechanisms for antibiotics depending on the working gas in the plasma system. The mechanisms can include the formation of ozone as most important active species, but also peroxy radicals that can lead further to formation of other oxidants and superoxide radicals which can hydrolyze to final products. It was concluded that ozone and hydroxyl radicals can affect the carboxyl group of the quinolone part and piperazinyl substituent and oxazinyl substituent.

3.2. Perfluorinated compounds

Perflurionated compounds are wide group of chemicals which are used in many industrial applications, as surfactants in non-stick cookware, packaging, textiles, firefighting foams production etc. Their characteristic C-F bonds are one of the strongest known bonds which makes those compounds environmentally persistent. In recent years, per- and polyfluoroalkyl substances (PFAS) had become of growing concern, especially long-chained compounds, such as perfluorooctanoic acid (PFOA) and perfluorooctane sulfonate (PFOS) due to their recalcitrant behavior (Lindstrom et al., 2011). In 2009, PFOS was listed as a Persistent Organic Pollutant (POP) in Stockholm Convention and by the United States Environmental Protection Agency (USEPA) PFOA listed as possible carcinogenic compound (Council, 2020; OECD, 2018; US EPA, 2016). Under the EU Water Framework Directive (2015/495 & 2015, 2015) PFOS and its derivatives are included as a priority hazardous substances with Environmental Quality Standard (AA-EQS) limit value of 0.65 ng/L in inland surface waters and 0.13 ng/L in seawater (ECHA, 2013; European Commission, 2000; Stockholm, 2019; UNEP, 2009). Therefore, global production of PFOA and PFOS started to be restricted or banned while new substances, as substitutes, were developed. In terms of wastewater treatment processes, many studies and practice had shown that most PFAS are highly recalcitrant to conventional wastewater treatment, but also to many AOPs. In the review paper by (Nzeribe et al., 2019) several physico-chemical processes, such as electrochemical oxidation, ultrasonication, plasma-based technology, advanced reduction processes (ARPs) for PFAS degradation have been compared and summarized. In general, the aim of those processes is complete mineralization of the compounds or at least degradation to non-toxic, biodegradable products. As shown, in terms of energy efficiency, treatment time and cost, plasma-based technology showed one of the most promising results and can be suitable for PFAS degradation.

Several authors have reported application of NTP for PFAS degradation. In (Yasuoka et al., 2011) degradation of PFOA and PFOS in solution by DC-plasma generated within gas bubbles was studied, using different gases. In oxygen plasma fluorine concentration after treatment was the highest while with helium plasma was the lowest. With argon plasma bubbles, the energy efficiency was the highest because of larger expansion along inner bubble surface, as authors suggested. Similar configuration was used in (Hayashi et al., 2015) where results showed degradation of PFOA after 3 hours of treatment, and 8 hours of treatment of PFOS. The evaluation of degradation efficiency was performed by measuring fluorine ions which were part of PFOA and PFOS molecules. Defluorination ratio of 94,5% confirmed almost complete degradation for PFOA. As for the PFOS, ratio of 70% can be explained with the existence of $SO_3H$ sulfo-group which makes PFOS less degradable. (Takeuchi et al., 2014) investigated degradation of PFOA using generated plasma inside oxygen bubbles. Initial concentration of 156 µM of PFOA was degraded after 2.5h treatment. Also, the authors investigated degradation of PFOA in a mix solution with perfluoroheptanoic acid (PFHpA), another perfluorocarboxylic acids (PFCA). In that case, rate constants decreased to about 80% compared to those in solutions containing single pollutant. Authors also pointed out that PFOA molecules behave differently in liquid-phase reactions then at the gas-liquid interface due to their surfactant characteristics which affects their degradation (Nozomi Takeuchi et al., 2011). (Stratton et al., 2017) applied laminar plasma jet with bubbling reactor which was used for degradation of PFOA and PFOS in solutions and in groundwater samples. PFOA was removed by 90% in a 30-minute plasma treatment with 76.5 W input power in high removal rate process. With input power of 4.1 W up to 25% of the pollutant was removed (high removal efficiency process). For the treatment of groundwater, the high removal efficiency process was used. Groundwater contained PFOA and several co-contaminants, such as PFOS, perfluorohexane sulfonate (PFHxS) and some non-fluorinated co-contaminants, trichloroethene and tetrachloroethene. To compare degradation efficiency results between the samples, the same process was used for treatment of prepared solution without other non-fluorinated co-contaminants and PFHxS. Degradation of PFOA had similar rate in both cases (within 2.5%) which can prove that there is no significant effect of other non-PFAS co-contaminants for PFOA degradation.

Possible degradation pathways for PFAS in a plasma-based water treatment have been proposed by several papers (Hayashi et al., 2015; Singh, Fernando, et al., 2019; Stratton et al., 2017; Takeuchi et al., 2014; Yasuoka et al., 2010). (Takeuchi et al., 2014) suggested the following degradation pathway: PFAS can adsorb on the gas-liquid interface since they have surfactant characteristics. Then, the C-C bond may break because of the interaction with electrons and ions but also as a result of a direct thermal degradation. Generated fluorocarbon radicals that are located in the bubbles may reduce and oxidize to H and O radicals, HF, $CO_2$ and CO gases. The authors indicated that there is more than one possible parallel pathway during degradation of PFAS, therefore, it is difficult to choose the exact degradation mechanism.

Decomposition of PFOA and PFOS was also investigated by (Hayashi et al., 2015). In the degradation process of PFOA, the by-products which are detected and measured were PFCAs ($C_nF_{2n+1}COOH$, $n$=1 to 6) with carboxyl group (COOH) like PFOA. Reaction formulas were suggested (Eqs.7-9) where $M^+$ stands for ions with the highest energy in the plasma (Hayashi et al., 2015).

$$C_nF_{2n+1}COO^- + M^+ \rightarrow C_nF_{2n+1}COO\cdot + M^+ + e^- \qquad (7)$$

$$C_nF_{2n+1}COO\cdot \rightarrow C_nF_{2n+1}\cdot + CO_2 \qquad (8)$$

$$C_nF_{2n+1}\cdot + 2H_2O \rightarrow C_{n-1}F_{2n-1}COO^- + 3H^+ + 2F^- + H\cdot \qquad (9)$$

The hypothesis is that the PFCA are present as negative ions on the surface, and they can collide with ions with the highest energy. Similar by-products may occur after degradation of PFOS in the treatment, but since PFOS has a sulfo-group ($SO_3H$), it is less likely decomposed than PFOA.

Also, the formation of PFOA as a by-product of PFOS is possible by reactions (Eqs.10-12) (Hayashi et al., 2015).

$$C_8F_{17}SO_3^- + M^+ \rightarrow C_8F_{17}SO_3\cdot + M^+ + e^- \qquad (10)$$

$$C_8F_{17}SO_3\cdot \rightarrow C_8F_{17}\cdot + SO_3\cdot \qquad (11)$$

$$C_8F_{17}\cdot + 2H_2O \rightarrow C_7F_{15}COO^- + 3H^+ + 2F^- + H \qquad (12)$$

In the study by (Yasuoka et al., 2010) it was reported that the decomposition of PFOA/PFOS may be significantly affected by the applied voltage polarity at the discharge electrode(s) in case of all working gases. Since PFOA/PFOS is an anion in the solution, concentrated close to bubble-water interface, in case of applied positive voltage to the powered electrode, it may react with some positive plasma species which come near the surface ($M^+$). Reactions involving these species were proposed (Yasuoka et al., 2010). The ionized PFOS molecules reacts with $M^+$ which leads to formed shorter carbon chains than PFOS and fluoride and sulfate ions are generated. The PFOA reacts in similar way as PFOS where fluorinated radical reacts with $H_2O$ yielding shorter-chain PFCA. Fluoride ions and carbon dioxide can be generated during PFCA degradation. Contrarily, if negative voltage is applied to the powered electrode, generation of oxygen and hydrogen occurs in the gas phase which may influence decomposition. Moreover, in this paper it was concluded that hydrated electrons are not significant for PFAS degradation. However, in the paper by (Stratton et al., 2017) where PFOA/PFOS degradation was studied, the hydrated electrons were reported as an important species in addition to argon ions and high-energy free electrons produced in the plasma. In this investigation, the shorter-chain by-products were quantified: PFHpA, perfluorohexanoic acid (PFHxA) and perfluoropentanoic acid (PFPnA). Although by-products were identified, it was shown that up to 10% of PFOA and PFOS is converted into shorter-chain PFAAs. Experiments with scavenger showed that hydroxyl and superoxide radicals play no significant role. Several by-products in plasma - based PFOA/PFOS solution treatments were also quantified by (Singh, Fernando, et al., 2019). PFOA, PFHxS and perfluorobutane sulfonate (PFBS) were found as by-products from PFOS degradation. Also, some fluoride ions, inorganic carbon, and smaller organic acids (trifluoroacetic acid, acetic acid, and formic acid) were identified with significant concentrations. The short chain PFCAs were detected as well, which suggests that some step-wise reduction of the parent compounds has occurred. Concentrations of all by-products increased in the first 60 minutes of treatment but by the end of experiment decreased. The authors suggested that the main species responsible for PFAS degradation are electrons from plasma/gas phase, aqueous electrons, and argon ions. Fig.2 from study shows proposed pathway for PFOA and PFOS degradation.

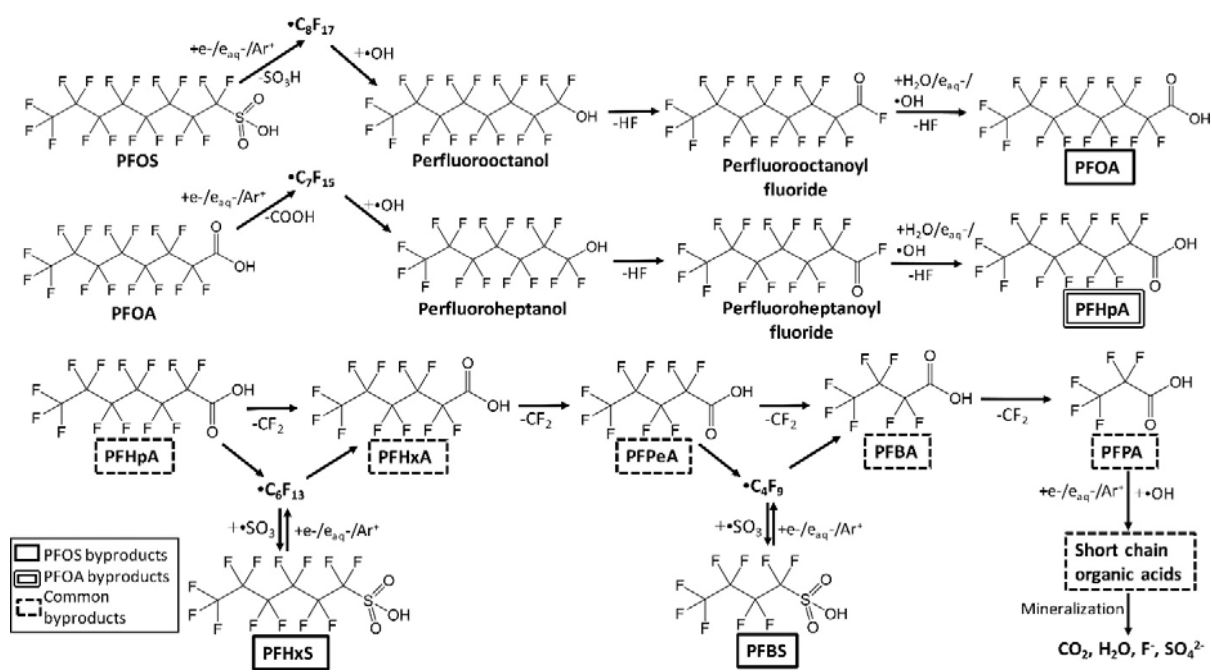

**Figure 2**. Proposed pathway for PFOA and PFOS degradation  Reprinted with permission from (Singh, Fernando, et al., 2019). Copyright (2019) American Chemical Society.

3.3. Pesticides

Pesticides are group of chemicals that are applied on the crops and soil against pest species. The huge variety of herbicides and insecticides are used worldwide in high quantities which makes them a group of contaminants with the highest potential to enter the environment. Recent studies have shown the potential of using NTP in liquid and gas-liquid environment for the remediation of pesticide manufacturing wastewater.

Plasma systems used in published studies for the pesticide removal are different types NTP, either discharges directly inside liquid or discharges in contact with liquid. Regarding the type of discharge, in most cases pulsed corona discharge and DBD with falling water film are used, some of them in combination with another AOPs (e.g., ozonation).

(Hijosa-Valsero et al., 2013) reported two different NTP reactors at atmospheric pressure, one operating as a thin-falling water film reactor and the other as conventional batch reactor, both producing discharge between dielectric barriers. The study compared degradation efficiencies for several pesticides: atrazine, lindane and chlorfenvinfos. For both reactors a first order degradation kinetics was proposed for all reactions. It was calculated that achieved relative removal rates for all pollutants in both reactors were around 70% after 5 minutes of treatment. In comparison to the batch DBD reactor above the liquid, energy efficiency (G-values, expressed in mol $J^{-1}$) were higher by one order of magnitude. That can be explained with the fact that in the coaxial reactor the electrode surface is larger allowing higher concomitantly treated volume of the sample. Similar configurations were reported in other studies (Hama Aziz et al., 2018; Vanraes et al., 2017; Wardenier, Vanraes, et al., 2019). (Vanraes et al., 2017) utilized a DBD reactor system with addition of activated carbon textile over the grounded inner electrode which caused better removal of all the pollutants. Indeed, depending on the molecular structure of a pollutant, plasma treatment can improve process of its degradation through adsorption. A DBD reactor with falling water film over activated carbon textile was also studied by (Wardenier, Vanraes, et al., 2019) for the degradation of pesticide mixture (atrazine, alachlor, diuron, dichlorvos and pentachlorophenol) where removal efficiencies between 57-88% were reported (56.9 % atrazine, 57.8% alachlor, 62.4% diuron, 69.3% pentachlorophenol, 70.2% dichlorvos). Many of the studies for the pesticide removal were performed by using one of the simplest plasma configurations - pulsed corona discharge above water (Bradu et al., 2017; Gerrity et al., 2010; Singh et al., 2016, 2017; Singh, Philip, et al., 2019). In a multiple needle-to-plane corona discharge reactor degradation of carbofuran from aqueous solution has been tested. Almost complete degradation occurred within 10 minutes of treatment and the degradation rate could be enhanced by increasing voltage and frequency (Singh et al., 2016). Similar configuration was reported in another study by (Singh et al., 2017) for the almost complete mineralization of 2,4-dichlorophenoxyacetic acid (2,4-D). For the same OMP (Bradu et al., 2017) generated NTP in a pulsed corona discharge combined with ozonation process. In this work, complete mineralization of 2,4-D after 30 minutes of treatment was demonstrated. Although the plasma treatment with corona discharge configurations have showed promising results, all of them worked with small treatment volume and in configurations where large-volume to plasma surface ratio have been preferred. However, a possible pilot-scale NTP device, with corona discharge in ambient air placed above the liquid surface was evaluated by (Gerrity et al., 2010). Experiments were conducted in different types of solution (wastewater and surface water). Promising removal efficiency was found for each type of water solution. Even though based on the results the oxidation is probably the dominant mechanism, more future studies are needed in the field of NTP for wastewater treatment applications with real samples type solution.

Several studies had proposed degradation pathways since the cognition of the mechanism and detection of intermediates is important step for pesticides plasma treatment optimization and practical applications. (Vanraes et al., 2015) detected ammelide and deethylatrazine as by-products in the degradation process of atrazine by a DBD in contact with liquid. The same intermediates were also reported by (Hijosa-Valsero et al., 2013). According to the literature, the main species responsible for the degradation of atrazine is hydroxyl radical and for plasma treatments this was confirmed in (Locke & Thagard, 2012). In some recent papers, decomposition mechanisms were also proposed for pesticide 2,4 – D, common used herbicide in agriculture (Bradu et al., 2017; Hama Aziz et al., 2018; Singh et al., 2017). In (Singh et al., 2017) the by-products were identified and several possible routes of breaking C-O bond on the benzene ring were described (Fig.3). Their conclusion again was that the main radical responsible for the degradation of the parent compound is *OH. The pathways were later confirmed by (Hama Aziz et al., 2018).

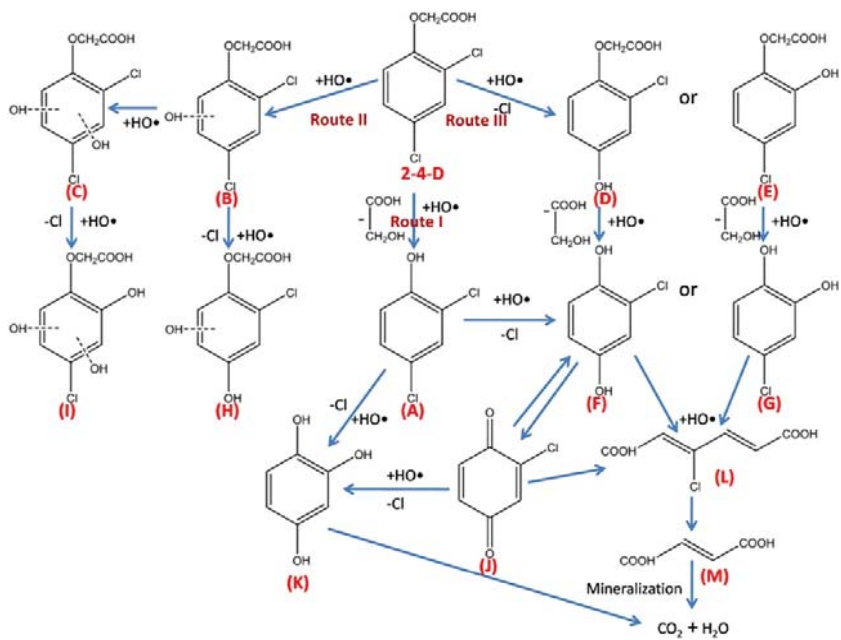

**Figure 3**. Degradation pathway of 2,4-D during plasma treatment. Reprinted with permission from (Singh et al., 2017) Copyright (2017) American Chemical Society.

In similar conditions, degradation of carbofuran was reported where seven intermediates were identified and reductive/oxidative pathways were proposed (Singh et al., 2016).

3.4. Phenols

Most common used models for comparing and testing plasma-based AOP are phenol and its derivatives (e.g. bisphenol A, nitrophenols, chlorophenols, hydroxyphenols). This group of chemicals are known to be used in many industrial applications as pesticides, disinfectants, antioxidants, etc. Their presence in ground, surface and drinking water has been of great concern due to the toxicity and estrogenic nature. As for the properties and oxidation pathways, they have been studied and well described in previous studies, but because of their very low biodegradability they still represent a challenge and more investigations of novel technologies for the improved removal of phenol and phenolic compounds are needed. According to the regulations of European Union Directives (2006/11/EC, 2013; 2008/105/EC, 2013), maximum allowable concentrations of phenols are 2,0 µg/L in inland surface waters and 0,3 µg/L for other surface waters.

In (Marotta et al., 2011) the degradation of phenol was investigated by application of DBD in air or in pure oxygen plasma above the liquid. The results showed that with the decreasing the air flow rate of phenol decomposition increases. The authors concluded that for phenol decomposition the most responsible are reactive species produced in the gas phase and then transferred into the solution. So, contact time between phenols and reactive species is of great importance since they have short lifetime. Regarding the influence of the electrode material on the treatment process, results has shown that after 4h treatment, the decomposition rate of phenol is about 3.2 times larger with stainless steel than Ni/Cr wires. The same experimental plasma configuration has been applied in (Marotta et al., 2012), where degradation rates of phenol in tap water and deionized water have been compared. The results shown significant increase of the degradation rate of phenol in tap water when compared to the one in deionized water. The authors suggested that possible reason is due to the specific effects of some other chemical species which can be found in tap water solution. Authors also investigated effects of the pH, bicarbonate ions, chlorine species, iron ions, etc. in the water on phenol degradation.

Bicarbonate ions had shown a positive effect on the degradation of phenols since they behave as a buffer and keep the pH of the solution around 7 during the plasma treatment. (Ceriani et al., 2018), is one of the first studies which reported complete mineralization of phenol, including mineralization of produced intermediates. It was found that after 30 min of treatment, phenol was completely removed, while mineralization of solution was achieved after 4h. In another experiment, with different initial concentration, after 5.5h complete mineralization was achieved. These results had shown that the initial concentration of OMP has effect on the degradation rate. On the other side, an electrical discharge generated directly inside water, based on a DC diaphragm discharge excited in a vapor bubble, was applied in (Lukes et al., 2013). About 50% degradation of phenol in $NaH_2PO_4$ solution after 50 min was achieved. Authors suggested that hydroxyl radicals formed by plasma led to the production of hydroxylated phenol products. Thus, hydroxyl radicals were found to be the main active components for the oxidation of phenol.

(Hijosa-Valsero et al., 2013) reported the degradation of 2,4 – dibromophenol in two different plasma reactors, one as a conventional batch (R1) reactor and other one as a coaxial thin- falling water film (R2), with a helium as a working gas. It was reported that relative degradation rate was around 98% in R1 after 5 minutes of treatment. Also, the kinetic constant (0.802 $min^{-1}$) was slightly faster in R1 and first-order degradation kinetics was proposed. (Hama Aziz et al., 2018) evaluated degradation of 2,4 – dichlorophenol (2,4 – DCP) in DBD planar falling water film reactor. In the study, comparable studies with ozonation and photocatalysis had also been reported. The authors concluded that in pure argon, after 15 min, complete degradation of 2,4 - DCP occurred in aqueous solution. The main reason is that hydrogen peroxide was produced under argon atmosphere in deionized water solution of 2,4 – DCP and high concentration of generated *OH radical. Several other papers (Banaschik et al., 2015; Chen et al., 2019; Iervolino et al., 2019) had also reported phenol degradation during plasma treatment and they evaluate the degradation effects depending on different types of discharge, working gas and the plasma reactor setup.

Several studies discussed possible mechanisms and pathways of the phenol and phenolic compounds degradation. In (Banaschik et al., 2015) high concentration of hydrogen peroxide was detected indicating that the production of hydroxyl radicals was significant. According to the authors, in Fig.4 possible degradation mechanism of phenol can be shown (Banaschik et al., 2015).

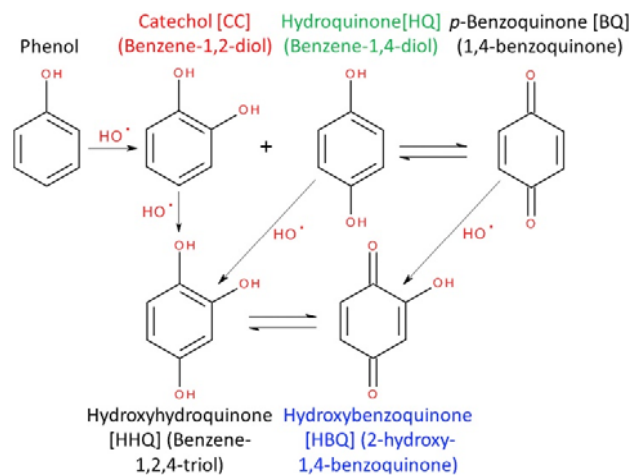

**Figure 4.** Possible reaction pathways of phenol with hydroxyl radicals. Reprinted with permission from (Banaschik et al., 2015). Copyright (2015) Elsevier.

Few phenol intermediates, such as benzoquinone (BQ), hydroxybenzoquinone (HBQ), catechol (CC) and hydroquinone (HQ) were detected. Results also shown that there is no trace of muconic acid or nitrated products of phenol. With that, it can be concluded that the ozone concentration was very low or absent. As for the intermediate, hydroxyhydroquinone (HHQ), the reaction product, HBQ was found, but not the HHQ himself. In (Lukes et al., 2013) phenol intermediates HQ, BQ, CC and HBQ were also detected. As for the other phenol products, the authors concluded that since in the diaphragm discharge was a present thermal mechanism it is possible that thermal decomposition occurred together with oxidative decomposition. (Ceriani et al., 2018) reported hydroxyl radical and ozone as the reactive species for the phenol degradation in miliQ water and formation of the intermediates: maleic acid, fumaric acid, *cis, cis*- and *trans, trans* - muconic acids, 1,2 – dihydroxybenzene, 1,4 – dihydroxybenzen, formic acid, acetic acid, oxalic acid. For all the intermediates, except maleic acid, it was found that they are no longer detected after 4h of plasma treatment. The results had shown that the concentration of formic and oxalic acids was increasing during the treatment while acetic acid reached $2.5 \times 10^{-5}$ M concentration and remained stable during 4h treatment. In (Chen et al., 2019) phenol and p-nitrophenol (PNP) were analysed by both HPLC and UV-VIS quantitative methods for different concentrations and different treatment times. With the increasing time, results had shown that concentrations of phenol and PNP decreased and decomposed into other intermediates under the influence of atmospheric pressure plasma jets (APPJs) in deionized water solution. Because of the presence of gas-liquid interfaces and the bubble motions, active species have been able to interact more with the phenol and PNP molecules. The degradation efficiency can also be enhanced by changing the pH value, according to the authors. There was no detailed study related to the degradation pathways, but the authors suggested the main reactions for both plasma configurations: inside and in contact with water. The analysis of the 2,4-DCP was performed after plasma treatment by high performance liquid chromatography (HPLC) in (Hama Aziz et al., 2018). It was found that chloride ion and other low chain anionic by-products occurred during the degradation process. Other by-products were identified such as oxalic acid, glyoxylic acid and glycolic acid. In most cases, the concentrations of the formed by-products increased during the beginning of degradation, but then decreased over time of the treatment, except oxalate ion which remain as a by-product even after completing the treatment.

3.5. Dyes

Dyes are OMP mainly used in textile industry, paper, plastic, and leather industry. The discharge of the dyes has been considered as great problem since most of these compounds show high resistance to the conventional wastewater treatment processes. Also, their toxicity and potential carcinogenicity raise concern for the aquatic ecosystem. Therefore, development of an effective wastewater treatment process for dye elimination had been in the focus of interest of many researchers. Recently, several plasma systems using different type of working gas (argon, air, oxygen) have been evaluated: DBD discharge with cylindrical configuration (He et al., 2018; Iervolino et al., 2019), pin to surface configuration (Kasih et al., 2019; Mitrović et al., 2020), flow reactor system (Yamada et al., 2020), corona discharge (Ruma et al., 2013), parallel plane type DBD reactor(Reddy & Subrahmanyam, 2012).

The degradation of methylene blue (MB) in a pulsed corona discharge, generated above the liquid surface in multiwire-plate was investigated by (Monica Magureanu et al., 2013). After about 10 minutes of plasma treatment, the MB solution decolorized rapidly and the formation of nitrate, formate, sulphate, and chlorine ions have been detected. Also, a colorimetric method was used to determine the presence of $H_2O_2$, main active species for the degradation of MB. It was found that during plasma exposure, concentration of $H_2O_2$ increased after 30 min treatment. Also, the presence of the $NO_3^-$ was detected, which can be prescribed to the air admixture and dissolved in water as the authors suggested since the oxygen was used as a working gas. The same compound was also investigated by (Kasih et al., 2019) and (Yamada et al., 2020). A pin to surface configuration plasma system at atmospheric pressure was used for decolorization of MB in water solution. After 120 min, decolorization efficiency was around 93%. In (Yamada et al., 2020), based on the results, decolorization efficiency of MB was around 90%. The experiments were performed using a gas-liquid pulsed discharge plasma with oxygen and argon as working gases. The concentrations of the active species were not tested but the results suggested that in the MB decolorization process, oxygen-based active species and *OH radicals were involved, especially in the case of argon gas. (Liu et al., 2016) suggested using atmospheric pressure DBD plasma system for the degradation of methyl orange (MO). The aim of this study was to evaluate new treatment method with glow discharge generated above the water surface. After 15 min, decolorization rate was 93 %. In (He et al., 2018), experiments were performed in the cylinder-like reactor, with a home-made direct current power source. It was concluded that the active species responsible for decolorization of MO are the short-lived ones instead of long-lived species such as $H_2O_2$. Nevertheless, the presence of $H_2O_2$ can affect the production of *OH radicals when in combination with Fenton process ($Fe_2^+$ ions). Several other authors evaluated different plasma configurations for dyes removal and up to this date, most of them concluded that the best efficiency is usually achieved when oxygen is used as a working gas. The details can be found in the reports (Iervolino et al., 2019; Jiang et al., 2012; Mitrović et al., 2020; Pankaj et al., 2017; Reddy & Subrahmanyam, 2012; Ruma et al., 2013; Sarangapani et al., 2017).

(Jiang et al., 2012) evaluated degradation mechanism for the decomposition of methyl orange and degradation efficiency depending on different initial concentrations. It was found that increasing initial concentration of MO decoloration efficiency (η) decreased through time. The authors suggested possible degradation pathway for the MO shown in Figure 5.

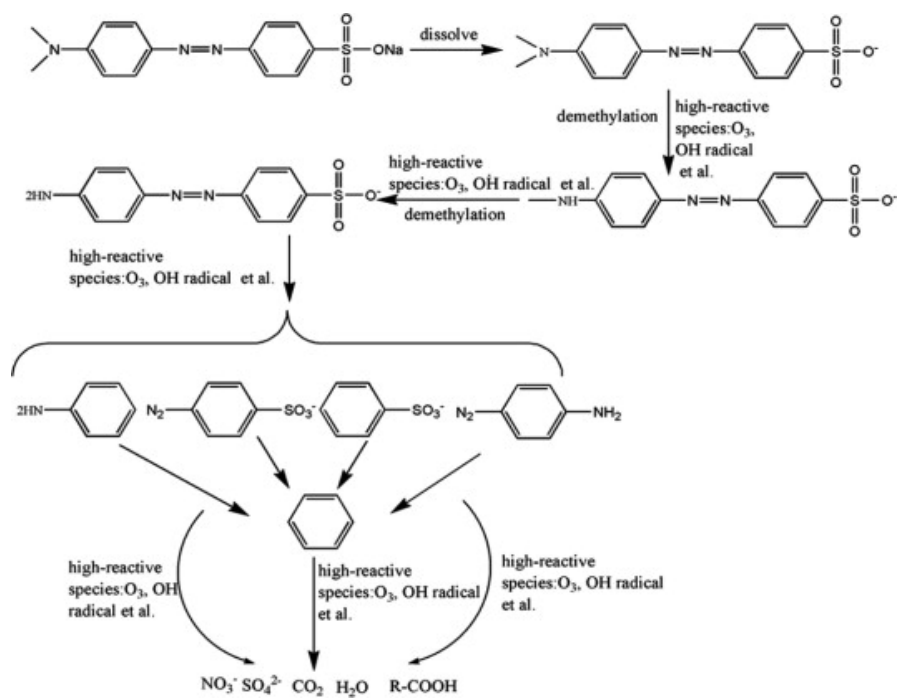

**Figure 5.** Proposed pathways for the degradation of MO using non-thermal plasma. Reprinted with permission from (Jiang et al., 2012). Copyright (2012) Elsevier.

Based on the dissociation energies (BDEs), authors suggested that the C-N-C bond BDE is the lowest and therefore breaking that bond results with two demethylation reactions which are the first ones taking place in plasma treatment. As for the C-N bond in the intermediate group ($C_6H_5$-$N_2C_6H_5$), the cleavage effect occurs because of the presence of the high reactive oxygen species, $O_3$ and *OH radical which are produced in a gas phase of the discharge. The authors also found the presence of $SO_4^{2-}$ ions and $NO_3^-$ which can be explained with possible mineralization of MO and the intermediates, into $CO_2$, $H_2O$, $SO_4^{2-}$ and $NO_3^-$. (Monica Magureanu et al., 2013) detected formic acid as an intermediate during MB plasma treatment. It is possible that the formation of acid resulted as a side chain oxidation. Also, small amounts of sulphate ($SO_4^{2-}$) were found. As in previous work, authors concluded a cleavage effect of the C-S bond in heterocyclic system. The generation of $H_2O_2$ has also been studied as important compound for the formation of *OH radicals. Comparing the concentration of $H_2O_2$ generated in water and in the MB solution through time, it was founded that higher concentration of $H_2O_2$ was generated in water. This can be related to the reactions between MB compound and its by-products with *OH radicals, influencing the less formation of $H_2O_2$. (Kasih et al., 2019) analyzed formation of MB functional groups through FTIR analysis. Some important MB functional groups were detected: -NH/-OH (overlapped vibration at 3417 $cm^{-1}$), -CH- (aromatic group at 2889 $cm^{-1}$), C-N (from amid II, at 1456 $cm^{-1}$), N-H (from amide III, 1112 $cm^{-1}$), C-N (from amide III, 852 $cm^{-1}$). From the spectrum, it can be seen that after 20 min of plasma treatment, a broad band in the range of 3000-3700 $cm^{-1}$ appears that can represent O-H stretching and a strong peak at 1643 $cm^{-1}$ as O-H-O bending scissors. The authors concluded that the high removal of 93% resulted after 120 min treatment.

Formation of reactive species during plasma treatment has been more investigated in some studies providing possible reaction sets for different working gases used (He et al., 2018; Jiang et al., 2012; Reddy & Subrahmanyam, 2012; Yamada et al., 2020). When in plasma system the oxygen was used as a working gas, as a result, there is a large generation of ozone. With dissociation of $H_2O$, $H_2O_2$ can be produced. Both ozone and $H_2O_2$ can be consumed for organic degradation. OH·, H·, O· and $HO_2$· radicals can be generated from various gases. *OH radicals can also be derived from $H_2O$ molecules which have been confirmed and evaluated by several authors as an important active species for the degradation (decolorization) of dyes.

4. Combining plasma technology and other advanced treatments

Several studies report application of NTP in a combination with other treatments. For example, (Mitrović et al., 2020), combined NTP with titanium dioxide ($TiO_2$) catalyst for the Reactive orange 16 (RO 16) dye degradation. The authors investigated the processes separately and in combination to compare energy efficiency of all systems. As a result, when NTP is enhanced with TiO$_2$ nanopowders, the degradation efficiency has improved since in the process the radicals are generated in the gas phase by the plasma and transported into the liquid, but also in the liquid on the surface of the catalyst. The influence of different flow rates and gas composition was also evaluated. In (Hama Aziz et al., 2017), ozonation, photocatalysis and DBD were compared for DCF and IBP in aqueous solution. The same reactor design was used in all experiments which allowed more valid comparison of the efficiency. For both pharmaceuticals, after DBD treatment, removal efficiency was higher than when treated with other methods, but also important note is that the degradation depended on the type of gas during treatment. Overall, the combination of two AOPs showed the most promising results. Similar conclusions were reported in (Hama Aziz et al., 2018) for pesticides. Other interesting reports can be found in (Vanraes et al., 2015), where degradation of atrazine was investigated by combination of NTP and nanofiber membrane and in (Vanraes et al., 2017), where NTP was combined with activated carbon. In both reports, results had shown that decomposition of atrazine was higher when NTP was used. Very detailed evaluation of physico-chemical processes for the PFAS treatment can be found in (Nzeribe et al., 2019). The authors investigated and compared efficiency such as cost, energy use, removal and proposed possible degradation pathway for several types of chemical oxidation and chemical reduction processes, ultrasonication and plasma technology. In general, as a proposed conclusion is that technologies which include persulfate and photochemical oxidation are less efficient than electrochemical oxidation technology. Based on the defluorination yield, energy use and cost, plasma treatment is one of the most efficient process for the PFAS treatment.

5. Outlook/conclusion

This paper gives a broad review on degradation mechanisms, pathways, removal efficiency of treatments by NTP for several groups of OMPs as well as the effects of different parameters on the degradation processes of pharmaceuticals, perfluorinated compounds, pesticides, phenols and phenolic compounds and dyes. For those OMPs which are known to be recalcitrant and difficult to remove with other types of AOP and conventional treatment, successful removal, and degradation up to 95% has been demonstrated with application of different types of discharges. In majority of cases plasma is generated in the air or other working gas or gas mixture above the liquid sample surface with the reactive species transferred from the gaseous to the liquid phase. Plasma sources used in these kinds of setups are different DBD geometries: coaxial, parallel plate, or thin falling water film, then (multi) pin-to-plane corona and streamer discharges operating mostly in pulsed mode. In fewer cases, the discharge is formed directly inside the liquid phase usually using pulsed power supplies. In some studies authors employed recirculation of the sample and showed that this results in enhancement of the decontamination in all cases. Most studies are done by studying an individual OMP present in the aqueous solution or, in some cases, in a synthetic mix solution with few other contaminants. It should be point out that these types of experimental set-ups are useful to study interaction between plasma-generated reactive species inside the liquid and pollutant molecules to better understand the degradation mechanisms for each contaminant. Once the mechanisms are identified, one can perceive the possibility of employment of the plasma technology for an efficient treatment of certain type of compound to be treated in polluted water. However, at this point little is known on how OMPs behave in real (waste)water samples when plasma is applied and what degradation pathways they would follow. As it can be seen from the review of the studies, depending on the type of OMP in water, different radical can play important role on its degradation. However, with other co-contaminants present in the water, the mechanism of the target OMP degradation can change easily since there is possibility that one or more of those co-contaminants have scavenger affect towards the radicals. An example for this is mixture of few different EC (pharmaceuticals and pesticides) in river water where presence of carbonates and bicarbonates, known as scavengers of *OH radicals, decreased degradation rate of EC. The results of most of the studies involving one or several contaminants had shown promising removal efficiency and even mineralization in some cases. Nevertheless, since most dominant reaction mechanisms are shown in "the best case" scenario, i.e. without the influence of co-contaminants, more research is needed with realistic samples of wastewater treated by plasma in order to apprehend the complete potential of the plasma technology. Because of this research gaps, further studies are recommended: (i) at laboratory scale – experiments using matrix solutions and real wastewater to understand optimal conditions and to elucidate degradation mechanisms for such complex matrix, (ii) at pilot scale – experiments using real matrix to investigate how to scale-up parameters for optimal energy/economic costs. According to the literature, most of the studies have been performed at laboratory scale and in such systems high degradation rates have been achieved. However, future studies should be also focused on the investigation of possibilities to scale-up the plasma systems and make application of the plasma technology at water volumes that are $10^4 - 10^6$ times higher than existing. Among other treatment parameters, this also affects the treatment duration; thus, the scaling-up of the systems should comprehend the changes in all related parameters. That is not a straightforward task and even though at laboratory scale plasma treatment showed numerous advantages compared to other AOP – such as no use of chemical reagents and better output of cost-efficiency, these assets should be also demonstrated at a large-scale in order to move the scientific advances towards new technology.

Moreover, the next important step in plasma degradation studies is further investigation degradation mechanisms and by-products formed both in the liquid phase and gas phase. In this research phase, it is very difficult to give proper general degradation mechanism for the organics due to the variations in given results by the research done up to this date. Possible degradation mechanisms and proper conclusions can be done per group of OMPs. Still, in some group of OMPs, those mechanisms can change based on the type of compound and working parameters, for example, by using different type of working gases, it is possible that different types of reactive species are produced, and possible reaction pathways can be changed. Therefore, the study of by-products is important. In recent years many new protocols have been reported for detection of those by-products after the AOP treatment and degradation of a parent compound. While some authors have investigated and studied the formation of by-products, but only few have evaluated and discussed the toxicity of those compounds. In some cases, they prove to be even more toxic and not biodegradable, for example PFAS. Therefore, this is an important step that should be investigated along with the decontamination process. Having said that, one possible way to resolve this issue is to focus the future studies towards reaching possible mineralization of treated compounds in plasma systems, especially for compounds where currently there is lack of findings of complete degradation and/or mineralization.

In terms of comparison between plasma and other AOPs, few studies were reported. Since the degradation efficiency depends on experimental conditions, i.e., many different treatment parameters (volume, initial concentration, characteristics of solution, reactor type, gas type, etc.), general comparison of this type is very difficult to obtain, and direct comparisons have been obtained only for specific cases. To have the reliable results for comparison, the combination of experimental conditions should be the same for all types of AOPs. Nevertheless, it was clearly shown when NTP is combined with other AOPs, better removal efficiency can be achieved, with reduced treatment time. Considering the necessity for scaling-up the plasma decontamination systems, combining plasma with another AOP seems a promising option to achieve large-scale reactors.

Acknowledgements

This work was supported by the European Union's Horizon 2020 research and innovation programme under the Marie Sklodowska-Curie grant agreement – MSCA-ITN-2018 (grant number 812880). ICRA researchers thank funding from CERCA program.